\documentclass[]{article}
\usepackage{packages}

\usepackage{arxiv}

\title{Insights from Game Theory into the Impact of Smart Balancing on Power System Stability}

\date{March 26, 2025}	

\author{ 
	\href{https://orcid.org/0009-0009-1660-9631}{\includegraphics[scale=0.06]{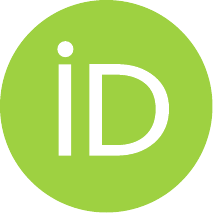}\hspace{1mm}Johannes Lips} \\
	Institute of Combustion and Power Plant Technology\\
	University of Stuttgart\\
	Stuttgart, Germany \\
	\texttt{johannes.lips@ifk.uni-stuttgart.de} 
	\And
	\href{https://orcid.org/0000-0002-0208-4100}{\includegraphics[scale=0.06]{orcid.pdf}\hspace{1mm}Hendrik Lens} \\
	Institute of Combustion and Power Plant Technology\\
	University of Stuttgart\\
	Stuttgart, Germany \\
}

\renewcommand{\shorttitle}{Insights from Game Theory into the Impact of Smart Balancing on Power System Stability}

\hypersetup{
pdftitle={Insights from Game Theory into the Impact of Smart Balancing on Power System Stability},
pdfauthor={J. Lips, H. Lens},
pdfkeywords={smart balancing, power system stability, game theory, experience-weighted attraction, frequency restoration reserve, imbalance settlement,  passive balancing},
}

\begin{document}
\maketitle

\begin{abstract}
Smart balancing, also called passive balancing, is the intentional introduction of active power schedule deviations by balance responsible parties (BRPs) to receive a remuneration through the imbalance settlement mechanism.
From a system perspective, smart balancing is meant to reduce the need for, and costs of, frequency restoration reserves (FRR), but it can also cause large oscillations in the FRR and jeopardize the system stability.
Using a dynamic control area model, this work defines a 2x2 game in which two BRPs can choose to perform smart balancing.
We study the impact of time delay, ramp rates, and pricing mechanisms on Nash equilibria and Experience-weighted Attraction (EWA) learning.
It is found that, even in an idealized setting, a significant fraction of games in a learned equilibrium results in an overreaction relative to the baseline disturbance, creating an imbalance in the opposite direction.
This suggests that the system stability risks are inherent to smart balancing and not a question of implementation.
Recommendations are given for implementation choices that can reduce (but not eliminate) the risk of overreactions.

\end{abstract}


\section{Introduction}\label{sec:intro}
\Glspl{tso} centrally coordinate the activation of \glspl{frr} to compensate for \gls{ca} imbalances and stabilize the power system.
The \gls{ca} imbalance, $P_\mathrm{ACE}$, with \glsunset{ace}\gls{ace} standing for \glsentrylong{ace}, is the sum of schedule deviations $P_b$ introduced by \glspl{brp} in that \gls{ca}. $P_b$ may for example be due to forecast errors, unplanned outages, or smart balancing.

In Europe, the \gls{tso} evaluates the incurred costs from \gls{frr} and the imbalance energy $E_b$ of the \glspl{brp} for each 15-minute timeframe, called \gls{isp}, and an imbalance settlement is made between each \gls{brp} and the \gls{tso} \citep{entsoeBalGL2018}.
Important for this settlement are the imbalance energy $E_b$, the \gls{frr} power (called \textit{balance delta} in \gls{nl}), $P_\mathrm{FRR}$, which is the sum of all \gls{frr} power in the \gls{ca} that was requested by the \gls{tso}, and the associated energy $E_\mathrm{FRR}$.
The energy amounts are calculated as
\begin{align} \label{eq:EintP}
    E_i = \int_\mathrm{ISP} P_{i}\left(t\right)\,\mathrm{d}t && i\in\left\{\mathrm{ACE,\,FRR,\,}b\right\}\,.
\end{align}
If the \gls{ca} was short ($E_\mathrm{ACE}<0$) during the \gls{isp}, positive \gls{frr} was activated, i.e., $E_\mathrm{FRR}>0$.
A \gls{brp} $b$ contributed to the \gls{ca} imbalance if it was short itself ($E_b<0$) and reduced the \gls{ca} imbalance if it was long ($E_b>0$).
This is reversed for \glspl{isp} during which the \gls{ca} was long ($E_\mathrm{ACE}>0$).

Two prices are defined for the imbalance settlement: $C^+$ is associated with the incurred costs for positive \gls{frr}, $C^-$ with the costs for negative \gls{frr}.
By definition, $C^-$ is smaller than zero if costs were made by the \gls{tso} for negative \gls{frr}. 
These prices can correspond to the maximum of the marginal costs of $P_\mathrm{FRR}$ during the \gls{isp}, as is the case in \gls{nl} \citep{koninkrijksrelatiesNetcodeElektriciteit}, or to a volume-weighted average of the marginal costs during the \gls{isp}, as is the case in \gls{de}.

With a \textit{single imbalance pricing} settlement mechanism, the imbalance price $C$ is
\begin{subnumcases} {C=\label{eq:C_single}}
C^+, & if $E_\mathrm{FRR}>0$,\\
C^-, & if $E_\mathrm{FRR}<0$.
\end{subnumcases}
The revenue made by \gls{brp} $b$ is 
\begin{equation} \label{eq:Pi_single}
    \pi_{b,\mathrm{DE}} = C \cdot E_b \,.
\end{equation}
If $\pi_b>0$, the \gls{tso} pays the \gls{brp} and vice versa.
Single imbalance pricing is used in \gls{de} and is the standard mechanism in Europe \citep{entsoeBalGL2018}.
Other options are \textit{dual imbalance pricing}, in which all schedule deviations of \glspl{brp} are penalized, and \textit{combined imbalance pricing}, which is used in \gls{nl} \citep{koninkrijksrelatiesNetcodeElektriciteit}, and uses single imbalance pricing unless there was a counter-activation of \gls{frr} with non-monotone $P_\mathrm{FRR}$ during an \gls{isp}, in which case dual pricing is used.
This can be formalized as 
\begin{subnumcases} {\pi_{b,\mathrm{NL}}=\label{eq:Pi_combined}}
C \cdot E_b, & if single pricing,\\
C^- \cdot E_b, & if dual pricing and $E_b>0$,\\
C^+ \cdot E_b, & if dual pricing and $E_b<0$.
\end{subnumcases}

Although \glspl{brp} should try to minimize their schedule deviation, single imbalance pricing provides an incentive for intentionally introducing nonzero $P_b$ in order to reduce the \gls{ca} imbalance and receive the corresponding imbalance remuneration.
In order to perform \textit{smart balancing}, \glspl{brp} evaluate their market position and the \gls{ca} imbalance situation.
Some \glspl{tso} (e.g., in \gls{nl} and Belgium) actively encourage smart balancing by publishing data on $P_\mathrm{FRR}$ and the associated costs in \gls{nrt}, improving transparency and reducing the uncertainty on $\pi_b$ \citep{elia1minute2019,SummaryBalanceDelta}.
Ideally, this would lead to reduced \gls{frr} costs, which was also suggested by the results of \citet{RoebenDiss2022} on the expected economic impact of publishing \gls{nrt} data in \gls{de}.

However, in November 2024, the Dutch \gls{tso} reported persisting $P_\mathrm{FRR}$ oscillations with amplitudes of up to \qty{1}{GW} due to the very large and fast response of \glspl{brp} to the published \gls{nrt} data.
The \gls{tso} noted that ``These oscillations~[\ldots] pose a risk to the system stability of the entire continent''~\citep{PowerImbalanceOscillations}.
They reacted by (temporarily) increasing the delay of the \gls{nrt} data from \qty{2}{min} to \qty{5}{min} and the dimensioned \gls{frr} volume, re-designing their \gls{afrr} controller, and limiting the ramp rate of assets to \qty{20}{\%/min} relative to their nominal capacity \citep{SummaryBalanceDelta}.
It is not certain that these measures will reduce the oscillations, because of the uncertainty about the interactions of the nonlinear pricing mechanism, the \gls{nrt} data, and the unknown control and decision structures used by \glspl{brp} for smart balancing.

A systematic study on the impact of the \gls{nrt} data and smart balancing on the system stability is therefore necessary, and has -- to the best of our knowledge -- never been done before.
From an economic perspective, game theory can be used to study the equilibria and learning dynamics for this problem.

\section{Game Description}\label{sec:game}
The following 2x2 game can be used to study the problem.
Two BRPs, referred to by $b \in \{1,2\}$, choose between performing smart balancing, $S_b = 1$, or not, $S_b = 0$, and do not have unintentional schedule deviations.
Starting from time $T_\mathrm{game}$, the \gls{nrt} data make clear that either BRP can make profit by performing smart balancing under the condition that the other BRP does not.
However, each BRP will encounter a loss if the other BRP also performs smart balancing because the combined reaction is larger than the baseline disturbance.

\begin{figure}
    \centering
    \includegraphics[width=0.7\columnwidth]{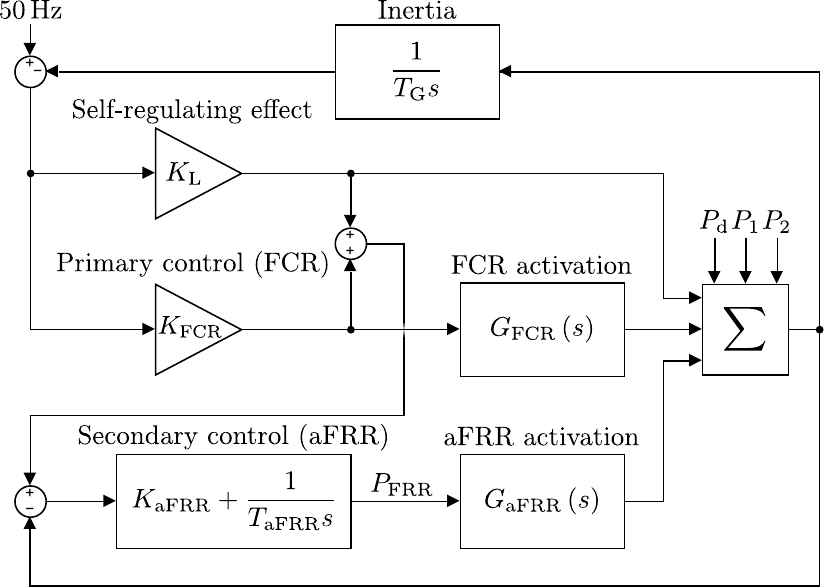}
    \caption{Single busbar model of a control area/control block.}
    \label{fig:modelCA}
\end{figure}

This setup can be dynamically simulated using the model presented in \cref{fig:modelCA}, which was parametrized to represent the German control block \citep{maucherControlAspectsInterzonal2022}.
The model contains the control and linearized activation dynamics of the \gls{fcr}, parameterized by $K_\mathrm{FCR}$ and transfer function $G_\mathrm{FCR}(s)$, respectively; the self-regulating effect $K_\mathrm{L}$; the \gls{afrr} control and activation, with parameters $K_\mathrm{aFRR}$, $T_\mathrm{aFRR}$, and transfer function $G_\mathrm{aFRR}(s)$; and the system inertia with time constant $T_\mathrm{G}$.
Since \gls{afrr} is the only \gls{frr} in the system, $P_\mathrm{FRR}$ is given by the output of the secondary (\gls{afrr}) controller.
The implementation is idealized because there is no exchange of power with other \glspl{ca} and the amount of primary control and self-regulating effect are exactly known.
In the rightmost block of the diagram, all physical power flows are summed.
This includes the power flows from smart balancing actions, $P_1$ and $P_2$, as well as the baseline disturbance $P_\mathrm{d}$, which contains the schedule deviations of all other \glspl{brp} in the control block.

\begin{figure}[t!]
    \centering
    \resizebox{0.8\columnwidth}{!}{%
%
%
\definecolor{mycolor1}{rgb}{0.50196,0.50196,0.50196}%
\begin{tikzpicture}

\begin{axis}[%
width=\columnwidth,
height=0.247\textwidth,
at={(0\columnwidth,0\textwidth)},
scale only axis,
xmin=-1,
xmax=30,
xtick={ 0,  5, 10, 15, 20, 25, 30},
xlabel style={font=\color{white!15!black}},
xlabel={Time $t$ [min]},
ymin=-210,
ymax=210,
ytick={-200, -150, -100,  -50,    0,   50,  100,  150,  200},
ylabel style={font=\color{white!15!black}},
ylabel={$P$ [MW]},
axis background/.style={fill=white},
xmajorgrids,
ymajorgrids,
legend style={at={(0.97,0.5)}, anchor=east, legend cell align=left, align=left, draw=white!15!black}
]
\addplot [color=mycolor1, line width=1.0pt, forget plot]
  table[row sep=crcr]{%
-1	0\\
30	0\\
};
\addplot [color=mycolor1, line width=1.0pt, forget plot]
  table[row sep=crcr]{%
15	-210\\
15	210\\
};
\addplot [color=black, dashdotted, line width=1.5pt]
  table[row sep=crcr]{%
-1	0\\
0	0\\
0	-200\\
30	-200\\
};
\addlegendentry{$P_\mathrm{d}$}

\addplot [color=blue, line width=1.5pt]
  table[row sep=crcr]{%
0	0\\
0.5	0\\
1	0\\
1.5	150\\
2	150\\
2.5	150\\
3	150\\
3.5	150\\
4	150\\
4.5	150\\
5	150\\
5.5	150\\
6	150\\
6.5	150\\
7	150\\
7.5	150\\
8	150\\
8.5	150\\
9	150\\
9.5	150\\
10	150\\
10.5	150\\
11	150\\
11.5	150\\
12	150\\
12.5	150\\
13	150\\
13.5	150\\
14	150\\
14.5	150\\
15	150\\
15.5	150\\
16	150\\
16.5	150\\
17	150\\
17.5	150\\
18	150\\
18.5	150\\
19	150\\
19.5	150\\
20	150\\
20.5	150\\
21	150\\
21.5	150\\
22	150\\
22.5	150\\
23	150\\
23.5	150\\
24	150\\
24.5	150\\
25	150\\
25.5	150\\
26	150\\
26.5	150\\
27	150\\
27.5	150\\
28	150\\
28.5	150\\
29	150\\
29.5	150\\
30	150\\
};
\addlegendentry{$P_b$ (400\%/min)}

\addplot [color=red, line width=1.0pt]
  table[row sep=crcr]{%
0	0\\
0.5	0\\
1	0\\
1.5	15\\
2	30\\
2.5	45\\
3	60\\
3.5	75\\
4	90\\
4.5	105\\
5	120\\
5.5	135\\
6	150\\
6.5	150\\
7	150\\
7.5	150\\
8	150\\
8.5	150\\
9	150\\
9.5	150\\
10	150\\
10.5	150\\
11	150\\
11.5	150\\
12	150\\
12.5	150\\
13	150\\
13.5	150\\
14	150\\
14.5	150\\
15	150\\
15.5	150\\
16	150\\
16.5	150\\
17	150\\
17.5	150\\
18	150\\
18.5	150\\
19	150\\
19.5	150\\
20	150\\
20.5	150\\
21	150\\
21.5	150\\
22	150\\
22.5	150\\
23	150\\
23.5	150\\
24	150\\
24.5	150\\
25	150\\
25.5	150\\
26	150\\
26.5	150\\
27	150\\
27.5	150\\
28	150\\
28.5	150\\
29	150\\
29.5	150\\
30	150\\
};
\addlegendentry{$P_b$ (20\%/min)}

\addplot [color=black!30!blue, line width=1.5pt, forget plot]
  table[row sep=crcr]{%
0	0\\
0.5	0\\
1	0\\
1.5	0\\
2	0\\
2.5	0\\
3	0\\
3.5	0\\
4	0\\
4.5	0\\
5	0\\
5.5	150\\
6	150\\
6.5	150\\
7	150\\
7.5	150\\
8	150\\
8.5	150\\
9	150\\
9.5	150\\
10	150\\
10.5	150\\
11	150\\
11.5	150\\
12	150\\
12.5	150\\
13	150\\
13.5	150\\
14	150\\
14.5	150\\
15	150\\
15.5	150\\
16	150\\
16.5	150\\
17	150\\
17.5	150\\
18	150\\
18.5	150\\
19	150\\
19.5	150\\
20	150\\
20.5	150\\
21	150\\
21.5	150\\
22	150\\
22.5	150\\
23	150\\
23.5	150\\
24	150\\
24.5	150\\
25	150\\
25.5	150\\
26	150\\
26.5	150\\
27	150\\
27.5	150\\
28	150\\
28.5	150\\
29	150\\
29.5	150\\
30	150\\
};
\addplot [color=black!30!red, line width=1.0pt, forget plot]
  table[row sep=crcr]{%
0	0\\
0.5	0\\
1	0\\
1.5	0\\
2	0\\
2.5	0\\
3	0\\
3.5	0\\
4	0\\
4.5	0\\
5	0\\
5.5	15\\
6	30\\
6.5	45\\
7	60\\
7.5	75\\
8	90\\
8.5	105\\
9	120\\
9.5	135\\
10	150\\
10.5	150\\
11	150\\
11.5	150\\
12	150\\
12.5	150\\
13	150\\
13.5	150\\
14	150\\
14.5	150\\
15	150\\
15.5	150\\
16	150\\
16.5	150\\
17	150\\
17.5	150\\
18	150\\
18.5	150\\
19	150\\
19.5	150\\
20	150\\
20.5	150\\
21	150\\
21.5	150\\
22	150\\
22.5	150\\
23	150\\
23.5	150\\
24	150\\
24.5	150\\
25	150\\
25.5	150\\
26	150\\
26.5	150\\
27	150\\
27.5	150\\
28	150\\
28.5	150\\
29	150\\
29.5	150\\
30	150\\
};
\addplot [color=black!50!blue, line width=1.5pt, forget plot]
  table[row sep=crcr]{%
0	0\\
0.5	0\\
1	0\\
1.5	0\\
2	0\\
2.5	0\\
3	0\\
3.5	0\\
4	0\\
4.5	0\\
5	0\\
5.5	0\\
6	0\\
6.5	0\\
7	0\\
7.5	0\\
8	0\\
8.5	0\\
9	0\\
9.5	0\\
10	0\\
10.5	150\\
11	150\\
11.5	150\\
12	150\\
12.5	150\\
13	150\\
13.5	150\\
14	150\\
14.5	150\\
15	150\\
15.5	150\\
16	150\\
16.5	150\\
17	150\\
17.5	150\\
18	150\\
18.5	150\\
19	150\\
19.5	150\\
20	150\\
20.5	150\\
21	150\\
21.5	150\\
22	150\\
22.5	150\\
23	150\\
23.5	150\\
24	150\\
24.5	150\\
25	150\\
25.5	150\\
26	150\\
26.5	150\\
27	150\\
27.5	150\\
28	150\\
28.5	150\\
29	150\\
29.5	150\\
30	150\\
};
\addplot [color=black!50!red, line width=1.0pt, forget plot]
  table[row sep=crcr]{%
0	0\\
0.5	0\\
1	0\\
1.5	0\\
2	0\\
2.5	0\\
3	0\\
3.5	0\\
4	0\\
4.5	0\\
5	0\\
5.5	0\\
6	0\\
6.5	0\\
7	0\\
7.5	0\\
8	0\\
8.5	0\\
9	0\\
9.5	0\\
10	0\\
10.5	15\\
11	30\\
11.5	45\\
12	60\\
12.5	75\\
13	90\\
13.5	105\\
14	120\\
14.5	135\\
15	150\\
15.5	150\\
16	150\\
16.5	150\\
17	150\\
17.5	150\\
18	150\\
18.5	150\\
19	150\\
19.5	150\\
20	150\\
20.5	150\\
21	150\\
21.5	150\\
22	150\\
22.5	150\\
23	150\\
23.5	150\\
24	150\\
24.5	150\\
25	150\\
25.5	150\\
26	150\\
26.5	150\\
27	150\\
27.5	150\\
28	150\\
28.5	150\\
29	150\\
29.5	150\\
30	150\\
};
\end{axis}
\end{tikzpicture}%
    }
    \caption{Disturbance $P_d$ due to a power plant outage at time $t=\qty{0}{min}$ and smart balancing reaction $P_b$ for $T_\mathrm{game} = 1, 5, \qty{10}{min}$ with fast and slow assets and $P_{b, \mathrm{max}} = \qty{150}{MW}$.}
    \label{fig:setupScenarios}
\end{figure}
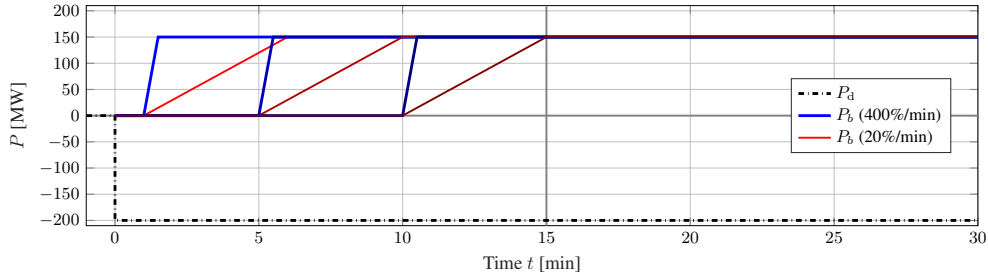

\begin{figure*}[tb!]
    \centering
    \resizebox{1\textwidth}{!}{%
    \input{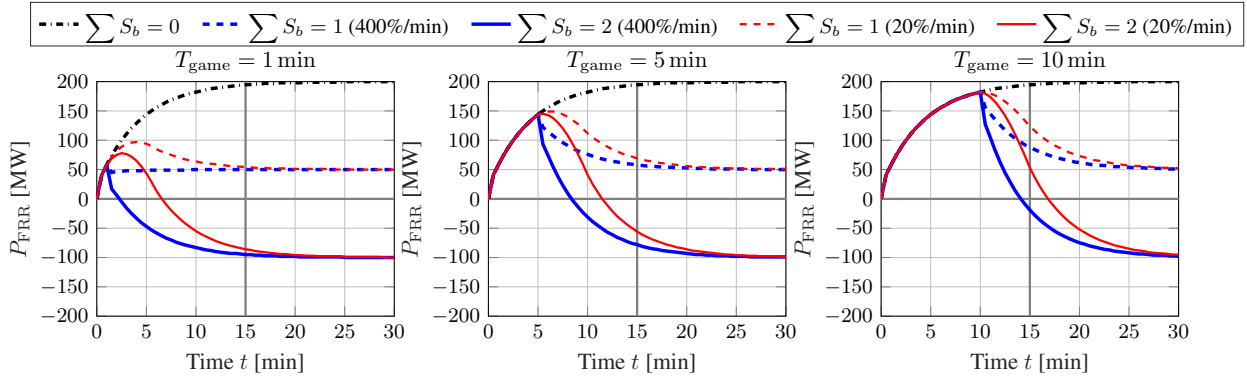}
    }
    \caption{Effect of smart balancing by one ($\sum S_b = 1$) or both ($\sum S_b = 2$) BRPs on the \gls{frr} power for symmetric scenarios, as defined in \cref{fig:setupScenarios}.}
    \label{fig:setupSimulations}
\end{figure*}

In \cref{fig:setupScenarios}, different smart balancing reactions to a disturbance, caused by a power plant outage of \qty{200}{MW} at time $t=\qty{0}{min}$, are shown.
The reactions vary in $T_\mathrm{game}$ and ramp rate $r$ ($| \dpdt |\leq r$) relative to the capacity $P_{b, \mathrm{max}} = \qty{150}{MW}$ of the asset that is used for smart balancing.
Note that the scenario in which a BRP introduces schedule deviations by trading on the intraday market (also called \textit{financial smart balancing}) is similar to $T_\mathrm{game}=\qty{1}{min}$.
The effect of smart balancing by one ($\sum S_b = 1$) or both ($\sum S_b = 2$) of the BRPs on $P_\mathrm{FRR}$ is shown in \cref{fig:setupSimulations} for games in which both BRPs use the same $T_\mathrm{game}$, $r$, and $P_{b, \mathrm{max}}$.
The setup clearly matches the one given at the beginning of this section: when a single BRP reacts (dashed lines), they reduce the need for \gls{frr}, when both react (solid lines), it causes an overreaction.
Multiple such overreactions can give rise to $P_\mathrm{FRR}$ oscillations.

Based on \cref{fig:setupSimulations}, it is possible to calculate the gains or losses that each BRP encounters over the two shown \glspl{isp} for different strategy profiles.
For calculating the payoffs, it is assumed that the marginal \gls{frr} prices scale linearly with $P_\mathrm{FRR}$. 
For the German imbalance pricing mechanism (\cref{eq:C_single,eq:Pi_single}, abbreviated as DE-pricing), this means that the positive imbalance price is calculated as
\begin{equation} \label{eq:CposDE}
    C^+_{\mathrm{DE}} = \int_\mathrm{ISP} \max\left(P_\mathrm{FRR}(t),0\right)\,\mathrm{d}t\,.
\end{equation}
The positive imbalance price for NL-pricing is given by
\begin{equation}\label{eq:CposNL}
    C^+_{\mathrm{NL}} = 0.25 \max_{t\in\mathrm{ISP}} P_\mathrm{FRR}(t)\,.
\end{equation}
The factor \qty{0.25}{(h)} scales the maximum power to a period of \qty{15}{min}, so that the prices are in the same range for \gls{de} and \gls{nl}.
For the calculation of $C^-$, $\min$ is used instead of $\max$.

The payoff table for this \textit{smart balancing game} is given in normal form in \cref{tab:payofftable} using variables $g_b$ for the gain obtained when BRP $b$ choses action $S_b = 1$, and $l_b$ for the loss suffered by BRP $b$ when both perform smart balancing.
For symmetric games, like the ones matching \cref{fig:setupSimulations}, $g$ and $l$ can be used:
\begin{subequations}\label{eq:symmetricGandL}
    \begin{align} 
        g_1 = g_2 &= g >0\,,\\
        l_1 = l_2 &= l >0\,.
    \end{align}
\end{subequations}
Values for $g$ and $l$, normalized to $\max(g)\!+\!\max(l)$, are given in \cref{tab:gainsNlosses} for different $T_\mathrm{game}$, $r$, and pricing mechanisms.
For the same $T_\mathrm{game}$ and $r$, $g$ and $l$ are higher with NL-pricing.
The ratio of $g$ to $g+l$ is systematically higher for DE-pricing.

\begin{table}
    \centering
    \caption{Payoff table for the smart balancing game.}
    \label{tab:payofftable}
    \setlength{\extrarowheight}{2pt}
    \begin{tabular}{cc|c|c|c}
      & \multicolumn{1}{c}{} & \multicolumn{2}{c}{BRP $2$}&\phantom{BRP1}\\
      & \multicolumn{1}{c}{} & \multicolumn{1}{c}{$S_2=0$}  & \multicolumn{1}{c}{$S_2=1$} &\\\cline{3-4}
      \multirow{2}*{BRP $1$}  & $S_1=0$ & $(0,0)$ & $(0,g_2)$ &\\\cline{3-4}
      & $S_1=1$ & $(g_1,0)$ & $(-l_1,-l_2)$& \\\cline{3-4}
    \end{tabular}
\end{table}

\begin{table}
    \centering
    \caption{Gains $g$ and losses $l$ for symmetric smart balancing games.}
    \label{tab:gainsNlosses}
    \begin{tabular}{llllllll}
\toprule
 $T_\mathrm{game}$ & $r$ & \multicolumn{3}{c}{DE-pricing} & \multicolumn{3}{c}{NL-pricing} \\
$\left[\mathrm{min}\right]$ & $\left[\mathrm{\%/min}\right]$ & $g$ & $l$ & $\frac{g}{g+l}$ & $g$ & $l$ & $\frac{g}{g+l}$ \\
\midrule
1 & 400 & 0.28 & 0.44 & 0.38 & 0.30 & 0.55 & 0.36 \\
1 & 20 & 0.31 & 0.35 & 0.47 & 0.37 & 0.48 & 0.44 \\
5 & 400 & 0.32 & 0.19 & 0.62 & 0.45 & 0.44 & 0.50 \\
5 & 20 & 0.32 & 0.15 & 0.67 & 0.42 & 0.37 & 0.53 \\
10 & 400 & 0.30 & 0.13 & 0.70 & 0.43 & 0.30 & 0.59 \\
10 & 20 & 0.27 & 0.11 & 0.71 & 0.45 & 0.19 & 0.71 \\
\bottomrule
\end{tabular}

\end{table}

\section{Analysis of the Smart Balancing Game}\label{sec:analysis}
\subsection{Nash Equilibrium Analysis}\label{sec:NE}

\cref{tab:gainsNlosses} holds the middle between the standard 2x2 games \textit{Battle of the Sexes} and \textit{Game of Chicken} \citep{Faliszewski2024}.
As in those games, there are two pure strategy \glspl{ne}: $(S_1,S_2) = (1,0)$ and $(S_1,S_2) = (0,1)$.
These strategy profiles can be rewritten as $(p_1^1,p_2^1) = (1,0)$ and $(p_1^1,p_2^1) = (0,1)$ using the notation $p_b^j$ for the probability that BRP $b$ uses strategy $j$.
With these strategy profiles, neither of the BRPs can improve their payoff through a one-sided change of strategy.
The additional mixed-strategy \gls{ne} can be written as
\begin{align} \label{eq:mixedNE}
&\phantom{\forall b \in \{1,2\}\,,} & p_b^1 = \frac{g}{g+l} && \forall b \in \{1,2\}\,.
\end{align}
Also for this $p_b^1$, there is a no-regret equilibrium.
Values for $g/(g+l)$ are listed in \cref{tab:gainsNlosses}.
The probability $p$ of an overreaction to the disturbance $P_\mathrm{d}$ is 
\begin{equation} \label{eq:overreact}
    p\left(\sum S_b = 2\right) = p_1^1\,p_2^1\,.
\end{equation}
For the pure strategy \glspl{ne}, $p_1^1p_2^1 = 0$.
For the mixed-strategy \gls{ne}, $p_1^1p_2^1$ can be as high as \qty{50}{\%} when $T_\mathrm{min}=\qty{10}{min}$ and $r = \qty{20}{\%/min}$.
This means that \qty{50}{\%} of the games played by BRPs using a mixed-strategy \gls{ne} would result in overreactions for this scenario.
Although this is a consequence of the optimal strategy of the BRPs, it is dangerous from a system stability perspective.

\subsection{Equilibrium Selection}\label{sec:NEselect}
To analyse which \gls{ne} would be selected when all players use rational criteria, the theory of equilibrium selection suggests the multilateral risk dominant \gls{ne}, which is the \gls{ne} with the smallest strategic risk,  as the solution for games in which there is complete information, i.e., all players know the complete payoff table \citep{harsanyiNewTheoryEquilibrium1995}.
The risk dominant \gls{ne} for the game in \cref{tab:payofftable} is 
\begin{subnumcases} {(S_1,S_2)=\label{eq:riskDominant}}
(1,0), & if $\frac{g_1}{g_1+l_1} > \frac{g_2}{g_2+l_2}$,\\
(0,1), & if $\frac{g_1}{g_1+l_1} < \frac{g_2}{g_2+l_2}$.
\end{subnumcases}

For the symmetric games in \cref{tab:gainsNlosses}, for which \cref{eq:symmetricGandL} holds, the conditions in \cref{eq:riskDominant} cannot be met and there is no single risk dominant \gls{ne}.
In such games, a correlated equilibrium, which requires previous communication between the players to alternate between the two pure strategy \gls{ne} is proposed as rational solution \citep{harsanyiNewTheoryEquilibrium1995}.
If this is not possible, the mixed-strategy \gls{ne} is a valid solution, but it is deemed undesirable, as it yields low payoffs and is nonpersistent \citep{harsanyiNewTheoryEquilibrium1995}.

\subsection{Experience-weighted Attraction Learning} \label{sec:ewa}
Learning algorithms offer an alternative approach to study how equilibria arise in games.
The \gls{ewa} model is often used, as it has been experimentally shown to describe the behaviour of real players relatively well, generalizes different learning rules, and only requires players to know their own payoffs \citep{camererEWALearningBilateral2002,pangalloTaxonomyLearningDynamics2022}.
In \gls{ewa}, players play the same game repeatedly and discretely update two state variables to learn a strategy based on the payoffs $\pi_b$ they received (reinforcement learning) or could have received given the other player's strategy (belief learning).
The states are the experience $N[k]$, with $k$ the discrete time at which the states are updated, and the attraction $A_b^{j}[k]$ that player $b$ has towards strategy $j$. 
Using $-b$ for the opponent(s) of player $b$, and $I$ for the identity function, \gls{ewa} is described by
\begin{subequations}\label{eq:ewa}
    \begin{align} 
        N[k] &= \left(1-\kappa\right)\left(1-\alpha\right)N[k-1]+1\,,\label{eq:ewa_experience}\\
        A_b^j[k] &= \frac{1}{N[k]}\Bigl(\left(1-\alpha\right)N[k-1]A_b^j[k-1]
                 + \bigl(\delta + \left(1-\delta\right)I\left(j,S_b[k]\right)\bigr)\,\pi_b\left(j,S_{-b}[k]\right) \Bigr)\,, \label{eq:ewa_attraction}\\
        p_b^j[k] &= \frac   {\exp\left( \beta A_b^j[k] \right)}
                            {\sum_{j'} \exp\left( \beta A_b^{j'}[k] \right)}\,,\label{eq:ewa_logit}
    \end{align}
\end{subequations}
with parameters $\delta, \alpha, \kappa$ and $\beta$.
$\delta$ ranges from 0 (reinforcement learning) to 1 (equal treatment of received payoffs and forgone payoffs).
The memory loss $\alpha$ ranges from 0 (infinite memory) to 1 (no memory).
The second depreciation parameter is $\kappa\in[0,1]$.
Finally, $\beta\in(0,\infty]$ determines the intensity of choice.
With $\beta = \infty$, $b$ always plays the strategy with the highest attraction.
Mixed strategies are played for finite $\beta$, which is a reasonable assumption, both for human behaviour and strategies learned by machine learning algorithms.

A smart balancing game with $10^4$ rounds is simulated to illustrate EWA in different scenarios, the parameters are chosen as $(\delta, \alpha, \kappa, \beta) = (0.25, 0.05, 1, 1)$.
These parameters indicate learning with a trade-off between reinforcement learning and belief learning, a small memory loss (valuing recent payoffs more) and mixed-strategies being played.
The initial experience $N[0]$ is $1$, and initial attractions $A_b^j[0]$ are sampled from the standard normal distribution, using the same initial attractions for different scenarios to enable comparisons.
Instead of updating \cref{eq:ewa} after each game, \cref{eq:ewa} is updated repeatedly after playing $100$ games with the same strategy, using the average payoffs obtained during those games in \cref{eq:ewa_attraction}.
This type of batch learning is realistic for the smart balancing game, and corresponds with BRPs updating their strategy on a regular basis, but not after every single \gls{isp}.

\begin{figure}
    \centering
    \resizebox{0.8\columnwidth}{!}{%
    \input{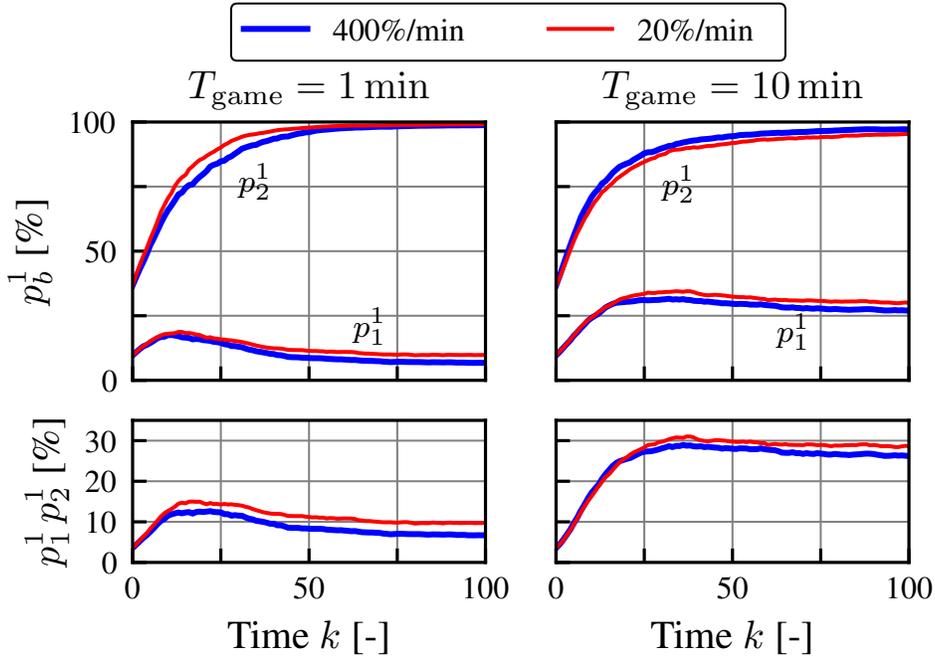}
    }
    \caption{Exemplary EWA learning dynamics with $(\delta, \alpha, \kappa, \beta) = (0.25, 0.05, 1, 1)$ for DE-pricing: (top) probability for smart balancing by BRP $b$, (bottom) probability of overreactions caused by smart balancing.}
    \label{fig:DEindividual}
\end{figure}

\begin{figure}
    \centering
    \resizebox{0.8\columnwidth}{!}{%
    \input{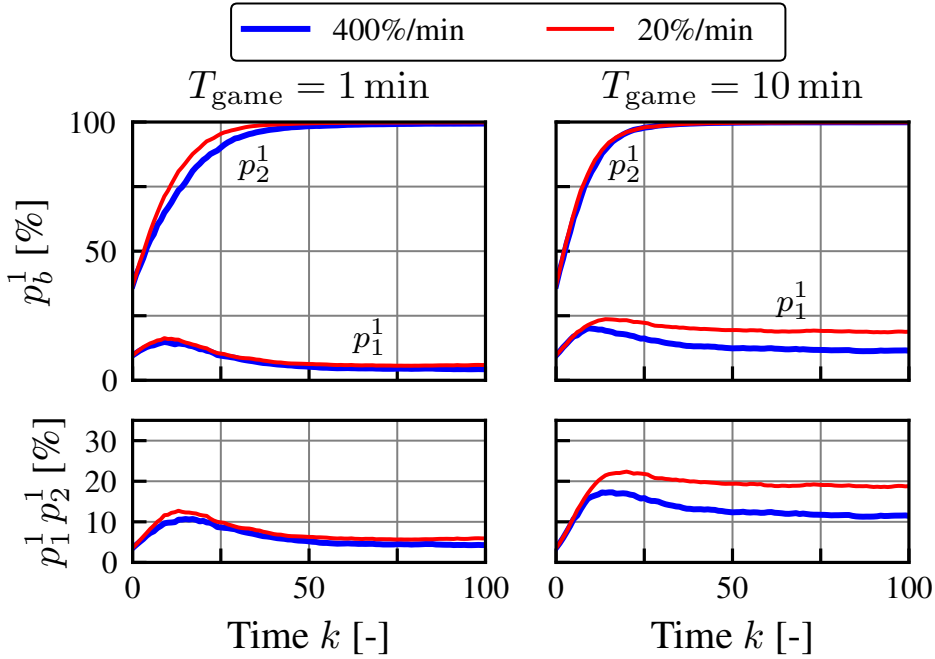}
    }
    \caption{Exemplary EWA learning dynamics with $(\delta, \alpha, \kappa, \beta) = (0.25, 0.05, 1, 1)$ for NL-pricing: (top) probability for smart balancing by BRP $b$, (bottom) probability of overreactions caused by smart balancing.}
    \label{fig:NLindividual}
\end{figure}

This is implemented building further on \citet{softwareFerrazEWA2022}.
The results for a single learning realisation are shown in \cref{fig:DEindividual} for symmetric games with the DE-pricing with $T_\mathrm{game} = \qty{1}{min}$ and \qty{10}{min}, and both studied $r$.
In the top figures, the probabilities $p_b^1$ that BRP $b$ performs smart balancing is shown, in the bottom figure the probability of overreactions \cref{eq:overreact} is shown.
In \cref{fig:NLindividual}, the results are given for NL-pricing.

The simulation converges to a pure stable \gls{fp} in which one of the BRPs (in this case BRP 2) almost always performs smart balancing, and the other BRP has a low (non-zero) probability of performing smart balancing.
The \gls{fp} is pure because it lies in the vicinity of a pure \gls{ne} (or at an alternative pure strategy profile) \citep{pangalloTaxonomyLearningDynamics2022}.
The probability of overreactions occurring when these strategies are used can be higher during the learning phase than for the final strategy, as is seen, e.g., in \cref{fig:NLindividual} for $T_\mathrm{game}=\qty{1}{min}$, where the probability of an overreaction is higher than \qty{10}{\%} for some $k<25$, but settles at less than \qty{5}{\%} at $k=100$.

With different learning parameters, the smart balancing game converges to a unique pure stable FP or NE, multiple pure stable FPs or NEs, or a unique mixed stable FP (a FP that does not lie in the vicinity of a pure NE) \citep{gallaComplexDynamicsLearning2013,pangalloTaxonomyLearningDynamics2022}.
Convergence to multiple FPs or NEs means that, depending on the initial attractions and experience, the dynamics converge to one of two stable solutions, as opposed to convergence to a unique stable solution.
Because the payoffs of the BRPs are positively correlated (both BRPs are unhappy with the solutions $(S_1,S_2) = (1,1)$, and to a lesser extent with $(S_1,S_2) = (0,0)$ -- they just don't agree on which of the NEs is best), chaotic solutions or limit cycles cannot occur for $k\rightarrow\infty$ \citep{gallaComplexDynamicsLearning2013}.
The convergence of the learning dynamics is good for the stability of the game, but convergence to FPs implies a non-zero probability of overreactions for $k\rightarrow\infty$.
In such a case (e.g., \cref{fig:DEindividual,fig:NLindividual}), smart balancing sometimes causes overreactions when the game is played repeatedly, and BRPs cannot be expected to change their strategy because of this, as they are satisfied with the overall payoff of the smart balancing.

In order to study the asymptotic behaviour of EWA learning algorithms, a series of simulations is performed for a multidimensional sweep through the learning parameter space.
It is assumed that the BRPs can play an infinite amount of games before updating \cref{eq:ewa}, so that they are aware of the opponent's mixed strategy.
This is a common assumption that can be formulated analytically \citep{pangalloTaxonomyLearningDynamics2022}).
The limit $k\rightarrow\infty$ is approximated with $k=100$ in the simulations.
The parameter ranges are $\delta\!\in\!\{0,0.25,0.5\}$, $\alpha\!\in\!\{0,0.05,0.1\}$, $\kappa\!\in\!\{0,0.5,1\}$, $\beta\!\in\!\{1,\infty\}$.
For each combination of parameters, 100 simulations are made with different initial attractions sampled from the standard normal distribution.
The parameter sweep is repeated for different pricing mechanisms, $T_\mathrm{game}$, and $r$.

\begin{figure}
    \centering
    \resizebox{0.8\columnwidth}{!}{%
    \input{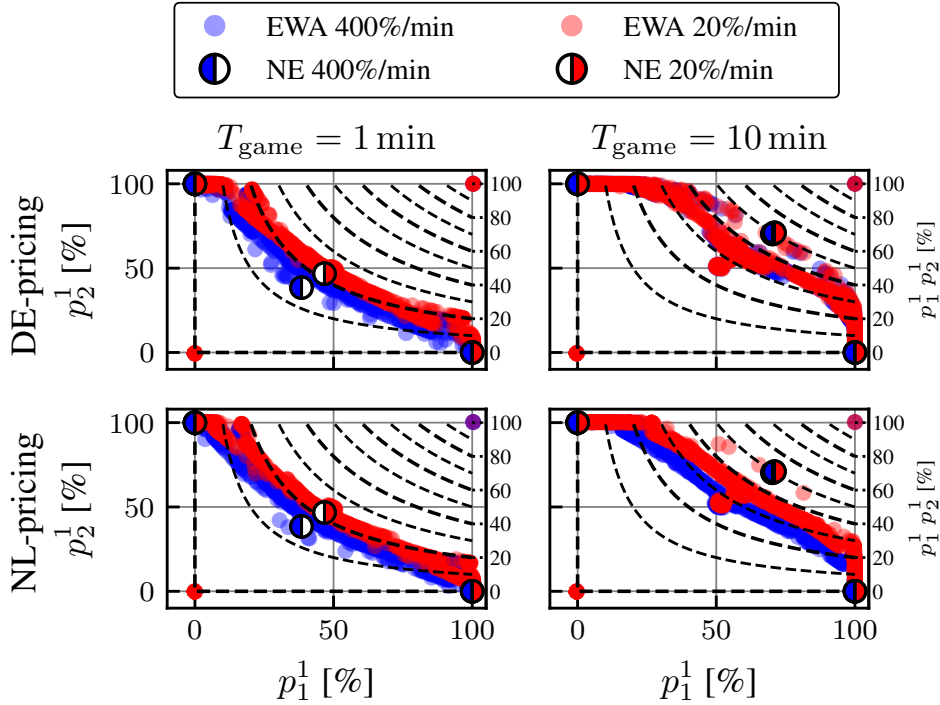}
    }
    \caption{NEs and EWA convergence points for the smart balancing game for different parameter combinations in four scenarios, showing probability $p_b^1$ that BRP $b$ performs smart balancing. 
    Isolines for $p_1^1p_2^1$ show the probability of overreactions in the power system introduced by smart balancing.}
    \label{fig:EWAfancyScatter}
\end{figure}

The results for $T_\mathrm{game} = \qty{1}{min}$ and \qty{10}{min}, for DE- and NL-pricing are shown in \cref{fig:EWAfancyScatter}.
The scatter plot shows the probabilities of $p_1^1$ and $p_2^1$ for different parameter combinations.
Additionally, the \glspl{ne} are shown for the different scenarios (see \cref{sec:NE}).
Finally, isolines for $p_1^1p_2^1$ show the probability of overreactions on baseline disturbance $P_d$ caused by smart balancing.
The solutions shown in \cref{fig:DEindividual,fig:NLindividual} are located near the upper left corner of the strategy spaces.

For all scenarios, simulations with convergence to pure and mixed FPs dominate the plot, and for certain parameter combinations, the steady-state solution results in overreactions for more than half of the ISPs when the game is played repeatedly ($p_1^1p_2^1>\qty{50}{\%}$).
The parameters that lead to pure strategy solutions ($\beta=\infty$) cannot be distinguished, as they all lie in the corners of the strategy space.
Games played at the beginning of the ISP (with small $T_\mathrm{game}$) have less probability of ending in an overreaction.
Faster assets lower the probability of overreactions as well, though the overall impact of the ramp rate of the assets is small.
The pricing mechanisms also influence the distribution of the FPs.

$T_\mathrm{game}$, $r$ and the pricing mechanism influence the outcome of the EWA learning because they affect $g$ and $l$.
A systematic analysis of the impact of $g$ and $l$ on the distribution of the FPs and the probability of overreactions can be used to understand how the payoff table (\cref{tab:payofftable}) can be constructed to minimize the probability of overreactions to disturbances.
An analytical solution of $p_{1}^{1}p_{2}^{1}$ in function of $g$, $l$ and the learning parameters does not exist, although progress is being made in creating a taxonomy of EWA in 2x2 games \citep{pangalloTaxonomyLearningDynamics2022}, where it is shown that both $l+g$ and $l-g$ play a role in the asymptotic behaviour of EWA.

In \cref{tab:Poverreaction}, the mean probability of overreactions and its standard deviation is given for the different symmetric scenarios studied throughout this paper, together with $l+g$ and $l-g$.
For EWA learning with $\beta\!=\!\infty$, convergence to pure NEs generally occurs, except when cumulative attractions and infinite memory are used ($\kappa\!=\!1$, $\alpha\!=\!1$), in which case players can `lock' into a certain strategy rapidly without exploring other possibilities (see \citet{camererEWALearningBilateral2002}, \citet[Fig. 2]{pangalloTaxonomyLearningDynamics2022}).
This results in a nonzero mean $p_{1}^{1}p_{2}^{1}$ and a significant standard deviation for the simulated parameter sweep.
For $\beta=1$, it is seen that $p_{1}^{1}p_{2}^{1}$ decreases with increasing $l+g$ and $l-g$.
In other words, when the absolute stakes are high (high $l+g$) and the potential losses are high compared with the potential gains (high $l-g$), the probability of both BRPs performing smart balancing at the same time, causing an overreaction, lowers.
That both $l+g$ and $l-g$ affect $p_{1}^{1}p_{2}^{1}$ is clear from the results with NL-pricing for $T_\mathrm{game}=\qty{5}{min}$, $r=\qty{400}{\%}$ and $T_\mathrm{game}=\qty{10}{min}$, $r=\qty{20}{\%}$.
These scenarios have the highest normalized value for $l+g$ (\num{0.89}) and the lowest value for $l-g$ (\num{-0.26}), respectively.
However, the scenario with $T_\mathrm{game}=\qty{5}{min}$ and $r=\qty{400}{\%}$ performs bad on $l-g$, and the second scenario performs good on $l+g$, so that they do not end up with the lowest, respectively highest, probability of overreactions.

\begin{table}
    \centering
    \caption{Mean overreaction probability $p_{1}^{1}p_{2}^{1}$ for a multidimensional parameter sweep of EWA dynamics, also showing $l\pm g$.}
    \label{tab:Poverreaction}
    \begin{tabular}{lllllll}
\toprule
 & $T_\mathrm{game}$ & $r$ & $l+g$ & $\;l-g$ & \multicolumn{2}{c}{$\text{mean}\,p_{1}^{1}p_{2}^{1}$ ($\text{std}\,p_{1}^{1}p_{2}^{1}$)}  \\
 & $\left[\mathrm{min}\right]$ & $\left[\mathrm{\%/min}\right]$ &  &  & \multicolumn{1}{l}{$\beta=1$} & \multicolumn{1}{l}{$\beta=\infty$} \\
\midrule
\parbox[t]{2mm}{\multirow{6}{*}{\rotatebox[origin=c]{90}{DE-pricing}}} & 1 & 400 & 0.72 & $\phantom{-}0.17$ & 15.6 (9.1) & 0.1 (3.6) \\
 & 1 & 20 & 0.66 & $\phantom{-}0.04$ & 17.5 (9.5) & 0.2 (4.3) \\
 & 5 & 400 & 0.51 & $-0.13$ & 22.1 (10.2) & 0.3 (5.1) \\
 & 5 & 20 & 0.47 & $-0.16$ & 23.9 (10.3) & 0.5 (7.3) \\
 & 10 & 400 & 0.42 & $-0.17$ & 24.9 (10.2) & 0.2 (4.3) \\
 & 10 & 20 & 0.39 & $-0.16$ & 25.5 (9.9) & 0.1 (3.3) \\
\midrule \parbox[t]{2mm}{\multirow{6}{*}{\rotatebox[origin=c]{90}{NL-pricing}}} & 1 & 400 & 0.85 & $\phantom{-}0.24$ & 14.3 (9.2) & 0.1 (3.3) \\
 & 1 & 20 & 0.86 & $\phantom{-}0.11$ & 15.4 (9.7) & 0.2 (4.7) \\
 & 5 & 400 & 0.89 &$ -0.01$ & 16.4 (10.3) & 0.2 (4.5) \\
 & 5 & 20 & 0.79 & $-0.05$ & 17.6 (10.3) & 0.4 (5.9) \\
 & 10 & 400 & 0.73 & $-0.13$ & 19.2 (10.5) & 0.3 (5.4) \\
 & 10 & 20 & 0.64 & $-0.26 $& 23.1 (11.2) & 0.1 (3.8) \\
\bottomrule
\end{tabular}

\end{table}

\section{Conclusions and Final Remarks}
Although smart balancing is meant to reduce the control area imbalance, it is possible that the combined smart balancing of multiple BRPs overcompensates the baseline disturbance $P_d$, causing an overreaction.
This is studied in a 2x2 game (\cref{tab:payofftable}) based on dynamic power system simulations, in which two BRPs can choose whether to perform smart balancing, receiving gain $g$ if they are the only one to perform smart balancing, and loss $l$ if they both perform smart balancing.
It is assumed that the BRPs know $g$ and $l$ at the time at which the game is played.
Different scenarios are used to analyse the effects of the different pricing mechanisms, as well as variations in the time at which the game takes place and the ramp rate of the used assets.

In this framework, the risk of smart balancing leading to overreactions to disturbances in the power system is shown to be inherent to the topology of the game: Neither equilibrium selection theory, nor \gls{ewa} learning, results in steady-state solutions without stochastic overreactions in repeated play.
Depending on $g$, $l$, and the learning dynamics, the probability of overreactions is up to \qty{50}{\%} for steady-state fixed points.
Learning dynamics can cause a temporary increase in this probability.
This can happen when a change in $g$ or $l$ triggers new learning dynamics, e.g., when changes are made to the imbalance pricing mechanism or the NRT data publication, or if new players enter the smart balancing game.

It is found that the probability of overreactions to disturbances can be reduced by increasing the stakes of the game (high $l+g$) and having a high potential loss relative to the potential gains (high $l-g$). 
The nonlinear effects in the imbalance pricing (the use of binary criteria, marginal prices and the multiplication of smart balancing energy with price) make it difficult to define general conditions that lead to the lowest risk of overreactions.
For the considered scenarios with a power plant outage at the beginning of the ISP, it is beneficial when BRPs play the game early in the ISP and use fast assets.
In these scenarios, the pricing mechanism used in the Netherlands leads to lower probability of overreactions than the German pricing mechanism.

Some of the measures taken by the Dutch TSO to reduce the negative effects of smart balancing are counterproductive from a game theoretical perspective.
The increase in delay for the publication of NRT data, ``done to increase the uncertainty of a result of a response based on the balance delta information'' \citep{SummaryBalanceDelta}, could result in games being played later in the ISP, increasing the probability of overreactions.
Clarity about potential payoffs (both $g$ and $l$) as early as possible in the ISP would be beneficial.
An additional positive measure would be a communication channel between BRPs, which would make correlated equilibrium solutions possible (\cref{sec:NEselect}), maximizing payoffs for the BRPs and minimizing the stability risk.
Further research could investigate whether this can be accomplished through the publication of additional \gls{nrt} data.

Because only the probability of overreactions is studied, the size and dynamics of the overreaction are not part of this study.
In \cref{fig:setupSimulations}, it is seen that a ramp rate limit of $r=\qty{20}{\%/min}$ on smart balancing assets slows down the dynamics of $P_\mathrm{FRR}$, achieving a positive effect on the system stability, even when this measure has a (limited) negative effect on the probability of overreactions.
Most findings of this study are expected to hold for an $n$ player game with multiple decision possibilities and asymmetric games, although for asymmetric games, equilibrium selection often results in pure NEs (see \cref{eq:riskDominant}).
For $n$ player games, the probability of $\left|E_\mathrm{FRR}\right|$ being larger with than without smart balancing can be investigated.
The effect of multiple consecutive decisions in sequential games (e.g., when \glspl{brp} notice an overreaction) can also be studied in future research.

\section*{Acknowledgment}
Many thanks to Marco Pangallo and Oussama Alaya for valuable conversations contributing to this paper.

\bibliographystyle{unsrtnat} 
\bibliography{zotero_for_GameTheory}

\begin{thebibliography}{13}
\providecommand{\natexlab}[1]{#1}
\providecommand{\url}[1]{\texttt{#1}}
\expandafter\ifx\csname urlstyle\endcsname\relax
  \providecommand{\doi}[1]{doi: #1}\else
  \providecommand{\doi}{doi: \begingroup \urlstyle{rm}\Url}\fi

\bibitem[{ENTSO-E}(2018)]{entsoeBalGL2018}
{ENTSO-E}.
\newblock Electricity {Balancing} in {Europe}, November 2018.
\newblock URL \url{https://eepublicdownloads.entsoe.eu/clean-documents/Network%20codes%20documents/NC%20EB/entso-e_balancing_in%20_europe_report_Nov2018_web.pdf}.

\bibitem[kon()]{koninkrijksrelatiesNetcodeElektriciteit}
Netcode elektriciteit.
\newblock URL \url{https://wetten.overheid.nl/BWBR0037940/2025-01-01#Hoofdstuk10}.

\bibitem[{Elia Transmission Belgium}(2019)]{elia1minute2019}
{Elia Transmission Belgium}.
\newblock End {User} {Documentation} “1-minute publications”, August 2019.
\newblock URL \url{www.elia.be/-/media/project/elia/elia-site/grid-data/balancing/20190827_end-user-documentation-elia1-minute-publications.pdf}.

\bibitem[{TenneT}(2024{\natexlab{a}})]{SummaryBalanceDelta}
{TenneT}.
\newblock Summary balance delta publication ({Follow}-up of the webinar on {November} 2, 2024), November 2024{\natexlab{a}}.
\newblock URL \url{https://tennet-drupal.s3.eu-central-1.amazonaws.com/default/2024-11/Summary%20balance%20delta%20publication%20webinar.pdf}.

\bibitem[Röben(2022)]{RoebenDiss2022}
Felix Röben.
\newblock \emph{Smart {Balancing} of {Electrical} {Power} {Transparent} {Real}-{Time} {Market} for {Cost}-{Efficient} {Power} {Balancing}}.
\newblock PhD thesis, Europa Universität Flensburg, 2022.
\newblock URL \url{https://d-nb.info/127953821X/34}.

\bibitem[{TenneT}(2024{\natexlab{b}})]{PowerImbalanceOscillations}
{TenneT}.
\newblock Power imbalance oscillations: {Balance} delta delay and additional measures ({TenneT} {Market} {News}), November 2024{\natexlab{b}}.
\newblock URL \url{https://www.tennet.eu/grids-and-markets/market-news}.

\bibitem[Maucher et~al.(2022)Maucher, Remppis, Schlipf, and Lens]{maucherControlAspectsInterzonal2022}
Philipp Maucher, Simon Remppis, Dominik Schlipf, and Hendrik Lens.
\newblock Control {{Aspects}} of the {{Interzonal Exchange}} of automatic {{Frequency Restoration Reserves}}.
\newblock \emph{IFAC-PapersOnLine}, 55\penalty0 (9):\penalty0 30--35, 2022.
\newblock ISSN 2405-8963.

\bibitem[Faliszewski et~al.(2024)Faliszewski, Rothe, and Rothe]{Faliszewski2024}
Piotr Faliszewski, Irene Rothe, and J{\"o}rg Rothe.
\newblock \emph{Noncooperative Game Theory}, pages 43--137.
\newblock Springer Nature Switzerland, 2024.
\newblock ISBN 978-3-031-60099-9.
\newblock \doi{10.1007/978-3-031-60099-9_2}.

\bibitem[Harsanyi(1995)]{harsanyiNewTheoryEquilibrium1995}
John~C. Harsanyi.
\newblock A new theory of equilibrium selection for games with complete information.
\newblock \emph{Games and Economic Behavior}, 8\penalty0 (1):\penalty0 91--122, 1995.
\newblock ISSN 08998256.
\newblock \doi{10.1016/S0899-8256(05)80018-1}.

\bibitem[Camerer et~al.(2002)Camerer, Hsia, and Ho]{camererEWALearningBilateral2002}
Colin~F. Camerer, David Hsia, and Teck-Hua Ho.
\newblock {{EWA Learning}} in {{Bilateral Call Markets}}.
\newblock In Rami Zwick and Amnon Rapoport, editors, \emph{Experimental {{Business Research}}}, pages 255--284. Springer US, Boston, MA, 2002.
\newblock ISBN 978-1-4757-5196-3.
\newblock \doi{10.1007/978-1-4757-5196-3_11}.

\bibitem[Pangallo et~al.(2022)Pangallo, Sanders, Galla, and Farmer]{pangalloTaxonomyLearningDynamics2022}
Marco Pangallo, James~B.T. Sanders, Tobias Galla, and J.~Doyne Farmer.
\newblock Towards a taxonomy of learning dynamics in 2 × 2 games.
\newblock \emph{Games and Economic Behavior}, 132:\penalty0 1--21, 2022.
\newblock ISSN 08998256.
\newblock \doi{10.1016/j.geb.2021.11.015}.

\bibitem[Ferraz(2022)]{softwareFerrazEWA2022}
Vinicius Ferraz.
\newblock Dynamic {Equilibria} {Prediction}: {Experience}-{Weighted} {Attraction} ({EWA}), 2022.
\newblock URL \url{https://doi.org/10.25937/q2xt-rj46}.
\newblock Publisher: CoMSES.Net.

\bibitem[Galla and Farmer(2013)]{gallaComplexDynamicsLearning2013}
Tobias Galla and J.~Doyne Farmer.
\newblock Complex dynamics in learning complicated games.
\newblock \emph{Proceedings of the National Academy of Sciences}, 110\penalty0 (4):\penalty0 1232--1236, 2013.
\newblock \doi{10.1073/pnas.1109672110}.

\end{thebibliography}

\end{document}